\documentclass[galaxies,article,accept,moreauthors,pdftex,10pt,a4paper]{Definitions/mdpi}
\usepackage{aas_macros}
\usepackage{pdflscape}
\firstpage{1}
\pubvolume{xx}
\issuenum{1}
\articlenumber{5}
\pubyear{2019}
\copyrightyear{2019}

\history{Received: 24 November 2018; Accepted: 14 January 2019; Published: 19 January 2019}
 \updates{yes}

\Title{H.E.S.S. Monitoring of PKS 2155-304 in 2015 and 2016}

\Author{Alicja Wierzcholska 
$^{1}$\orcidA{}*, Michael Zacharias $^{2,3}$\orcidB{}, Felix Jankowsky $^{4}$, Stefan Wagner $^{4}$ and for the H.E.S.S. Collaboration} 

\AuthorNames{Alicja Wierzcholska, Michael Zacharias, Felix Jankowsky and Stefan Wagner}

\address{%
$^{1}$ \quad Institute of Nuclear Physics, Polish Academy of Sciences, PL-31342 Krakow, Poland\\
$^{2}$ \quad Institut  für  Theoretische  Physik,  Lehrstuhl  IV:  Weltraum und Astrophysik, Ruhr-Universität Bochum, \mbox{D 44780 Bochum,} Germany; mzacharias.phys@gmail.com\\
$^{3}$ \quad Centre for Space Science, North-West University, Potchefstroom, 2520, South Africa\\
$^{4}$ \quad Landessternwarte, Universität Heidelberg,  
Königstuhl, D-69117 Heidelberg, Germany; jankowsky@lsw.uni-heidelberg.de~(F.J.); swagner@lsw.uni-heidelberg.de (S.W.)}

\corres{Correspondence: contact.hess@hess-experiment.eu or alicja.wierzcholska@ifj.edu.pl
}

\abstract{PKS\,2155-304 is one of the brightest blazar located in Southern Hemisphere, monitored with H.E.S.S. since the first light of the experiment.
Here we report multiwavelength monitoring observations collected during the period of 2015--2016 with H.E.S.S., \emph{Fermi}-LAT, Swift-XRT, Swift-UVOT, and ATOM.
Two years of multiwavelength data with very good temporal coverage allowed to characterize broadband emission observed from the region of PKS\,2155-304 and study potential multifrequency correlations.
During the period of monitoring, PKS\,2155-304 revealed complex multiwavelength variability with two outbursts characterized by completely different multiband properties.
The 2015  activity of the blazar is characterized by a flare observed at all wavelengths studied. The broadband emission observed during the outburst is well correlated without any time lags.
Contrary to 2015, in 2016, only orphan outburst in the optical and ultraviolet wavelengths was observed. Such an orphan activity is reported for the first time for the blazar PKS\,2155-304.} 

\begin{document}

\section{Introduction}
BL Lacertae (BL Lac) type objects and flat spectrum radio quasars (FSRQs) constitute a class of blazars.
According to the unified scheme of active galactic nuclei (AGNs), blazars are AGNs that are characterized by jets pointing at small angles with respect to the line of sight \citep{Urry95}.
The radiative energy output from these sources is highly dominated with Doppler-boosted nonthermal emission (e.g., \citep[]{Begelman84}).

Blazars are sources that are known for their temporal variability, which can be observed at different wavelengths and on various timescales (see e.g., \citep[]{Wagner09, Wierzcholska15L, Siejkowski17, Wierzcholska07full}).
The~spectral energy distribution (SED) of blazars, in the $\nu$-$\nu$F$_{\nu}$ representation, is characterized by a double-humped structure.
The~low-energy bump in the SED is usually attributed to the synchrotron radiation of ultra-relativistic electrons accelerated within the jets.
The high-energy bump is generally discussed in the context of different scenarios.
In the most popular leptonic explanation, the  second  bump  is  interpreted  as  the inverse  Compton  radiation from  the  same  population  of  relativistic  electrons  scattering the synchrotron photons (so-called  Synchrotron Self Compton models; for details see e.g. \cite{Maraschi92, Kirk98}) or external  photons  from  the accretion
disc,  broad  lines  region  or  dust  (EC,  External  Compton model; for details see e.g.,~\cite{Dermer92, Sikora94}).
As alternative hadronic scenarios have been proposed (e.g., \citep[]{Aharonian2000, Atoyan2003, bttcher13}).

The frequency of low energy peak in the blazars' SED has been used as a parameter to divide these objects into subclasses of high-, intermediate-
and low-energy peaked BL Lac objects: HBL, IBL, LBL, respectively \citep[][]{Fossati98, Abdo10}.
In the case of HBL blazars, the synchrotron peak is located in the X-ray range, for IBL objects in the optical-UV range, and for LBL type blazars in the infrared regime \citep{Abdo10}.

The blazar PKS\,2155-304 (redshift $z = 0.116$) is one of the brightest HBL type object, located in the Southern Hemisphere. Its strongly variable non-thermal emission was observed in different energy bands for already more than 40 years.
The object was discovered in the radio frequencies as part of the Parkes survey \citep{Shimmins74},  and later identified as a BL Lac type blazar \citep{Hewitt80}.

Since 2004, in the very high energy $\gamma$-ray regime, PKS\,2155-304 was frequently observed with the  H.E.S.S. telescopes, and this resulted in the discovery of different activity states including low flux level and flaring activity of the object.

In July 2006 ,an exceptional $\gamma$-ray flare was detected with H.E.S.S. four telescopes.
The outburst was characterized in the very high energy regime (VHE, E $>$ 100\,GeV)  with fluxes  40 times the average flux, and flux doubling timescales of the order of a few minutes.
Such an ultrafast variability requires very large Doppler factors of the jets and suggests that the observed variability is not directly connected to the central black hole instantaneous activity \citep{2155_flare}.

The second flaring event detected in PKS\,2155-304 in July 2006 was observed during simultaneous MWL observations with H.E.S.S. and Chandra \citep{2155_chandra}.
It revealed a strong correlation both in flux and spectrum, with no evidence of lags.
However, during the decaying phase of the flare cubic relation between fluxes for X-ray and $\gamma$-ray fluxes has been observed.
No universal optical–VHE correlations were found in the 2006 flaring data.
Analysis of longterm VHE monitoring data including H.E.S.S. observations collected between 2004 and 2012 and \emph{Fermi}-LAT observations taken between 2008 and 2012 showed that emission observed in the quiescent state of PKS\,2155-304 is consistent with a log-normal spectral behavior  \citep{lognorm}.

\section{Materials and Methods}
\vspace{-6pt}
\subsection{VHE $\gamma$-Ray Observations with H.E.S.S.}
The High Energy Stereoscopic System (H.E.S.S.) is an array of five Imaging Atmospheric Cherenkov Telescopes (IACTs), located  in the Khomas Highland in Namibia.
The observations are performed in the regime of  very-high-energy  $\gamma$ rays (VHE, $E>100$\,GeV) (for details see \citep[]{Aharonian2006_crab}).
Since 2012, the system consists of  four 12\,m telescopes, each with a mirror area of  108\,m$^2$,  a fifth, larger telescope with a mirror area of  614\,m$^2$.

The blazar PKS\,2155-304 was observed with the H.E.S.S. telescopes in 2015 and 2016 in 91 and 40\,runs, respectively.
This corresponds to 36 and 16.5\,h of observations.
The analysis was performed using ImPACT analysis chain \citep{Parsons14}  with the Loose Cuts configuration \citep{Aharonian2006_crab}.
Similar results as those presented in the paper were obtained using an independent analysis chain (the Model Analysis,~\citep[]{Naurois09}).

\subsection{HE $\gamma$-Ray Observations with Fermi-LAT}
High energy $\gamma$-ray data (HE,  E $>$ 100\,GeV) collected with the \emph{Fermi}-LAT in the period of  1 January 2015, and  31 December 2016, have been analyzed using standard Fermi Science Tools (version v10r0p5) with P8R2$\_$SOURCE$\_$V6 instrument response functions   \cite{Atwood13}.
The analysis was performed using all events in the energy range from 100 MeV to 300 GeV and zenith angles were limited to 90 degrees.
The region of interest was constrained to have a size of 10\,degree and be centered on the source.
As~an analysis method binned maximum-likelihood was applied with the Galactic diffuse background modeled using the \verb|gll_iem_v06| map cube, and the extragalactic diffuse and residual instrument backgrounds modeled jointly using the \verb|iso_P8R2_SOURCE_V6_v06| template \citep{Mattox96}.

\pagebreak
\subsection{X-ray and Ultraviolet Observations with Swift-XRT and Swift-UVOT}
X-ray, optical and ultraviolet observations collected with Swift-XRT and Swift-UVOT, respectively with ObsID  00049686003-00049686005 were analysis.
These ObsIDs correspond to all data collected in 2015--2016. 
In the case of X-ray observations, the analysis was performed using \verb|HEASOFT| v.6.23 and for spectral fitting  \verb|XSPEC| v.12.9.1 was used.
All data were binned in order to have at least 20 counts per bin and fitted using a single power-law model with a Galactic absorption value of 1.52 $\times$ 
 10$^{20}$\,cm$^{-2}$ \citep{Kalberla2005}  set as a fixed parameter.

In the case of optical/ultraviolet observations,  instrumental magnitudes were calculated using \verb|uvotsource|.
A region of interest and was defined as a circle with a radius of 5\,arcsec.
Same size region, located close to the source region, but being uncontaminated with any signal from the nearby sources,  was used in the case of background definition.
The~flux conversion factors were taken from \cite{Poole08}.
The~UVOT data were corrected for the influence of the Galactic extinction based on the model from \cite{Schlegel98} with the most recent recalibration by \cite{Schlafly},
using the Extinction Calculator from NED (NASA/IPAC EXTRAGALACTIC DATABASE).

\subsection{Optical Observations with ATOM}
In 2015--2016 PKS\,2155-304 was also monitored with the 75\,cm Automatic Telescope for Optical Monitoring (ATOM) located in Namibia at the H.E.S.S. site in $B$ and $R$ bands \citep{Hauser}.
The magnitude of each observation was derived using differential photometry using six secondary standard stars located in the field of view of PKS\,2155-304.

The magnitudes have been corrected against the Galactic extinction.
For the spectral studies, ATOM data were also corrected for the contribution of the host galaxy, using the template of an elliptical galaxy provided by \cite{Fukugita95} and observations in Gunn filter $i$  \cite{Falomo91}, with an assumed de Vaucouleurs profile of the starlight.

\section{Results}
\vspace{-6pt}
\subsection{Longterm Variability}
\begin{figure}[H]
\centering
\includegraphics[width=0.98\textwidth]{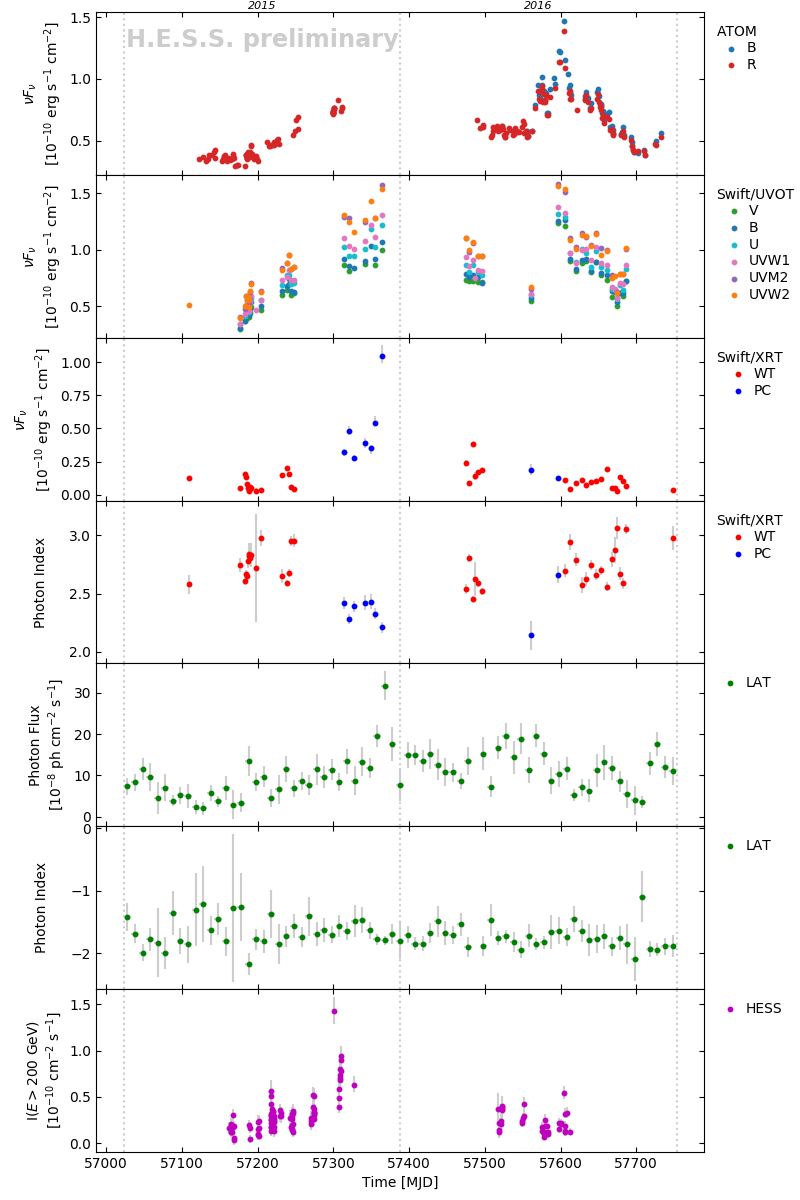}
\caption{Multiwavelength light curve of PKS\,2155-304 presenting data collected in 2015--2016 with ATOM, Swift-UVOT, Swift-XRT, Fermi-LAT and H.E.S.S.}
\label{fig:lc2155}
\end{figure}

The longterm multiwavelength light curves presenting observations of PKS\,2155-304 performed in 2015--2016 is shown in Figure\,\ref{fig:lc2155}.
From top to bottom panels present: optical fluxes in B and R bands as observed with ATOM, optical and ultraviolet observations taken with Swift-UVOT, Swift-XRT flux and photon index,  Fermi-LAT flux and photon index, and flux above 200\,GeV observed with H.E.S.S.
Fermi-LAT data are binned into 10-day intervals.
In all other cases, 1-day bins are used.
Significant variability of PKS\,2155-304 is observed at all wavelengths.
However, the trends visible in both years are different.
In 2015, an increase of the flux observed at all wavelengths has been observed.
In the VHE $\gamma$-ray regime the highest point in the light curve correspond to the flux of about 1.5 $\times$ 10$^{-10}$\,cm$^{-2}$\,s$^{-1}$, which is an equivalent to 3 times the flux observed from the direction of PKS\,2155-304 during its quiescent state as defined in \citep{lognorm}.
The larger increasing activity corresponds to constant \emph{Fermi}-LAT photon index values and hardening of the X-ray photon flux observed with Swift-XRT.
In 2016, there~was also flaring activity observed in multiwavelength observations of PKS\,2155-304, but in this case, the flare was detected only in an optical and ultraviolet regime.
No counterpart either in X rays or in $\gamma$ rays has been observed.

The multiwavelength data provide good coverage of both bumps in the SED of PKS\,2155-304 both for 2015 and for 2016.
To quantify the broadband variability, the fractional variability amplitude is used.
It is defined following \cite{Vaughan2003} as:
\begin{equation}
  F_{var} = \frac{\sqrt{S^{2} - \sigma^{2}_{err}}}{<F>},
\end{equation}
 where $<F>$ is an average flux, $S^{2}$ is the variance,  and $\sigma^{2}_{err}$ is the square mean error.
The resulting F$_{var}$ values for different instruments are listed in Table\,\ref{table:fvar2155}.
The comparison of yearly variability quantified with  F$_{var}$, shows that the source was more variable in 2015 than  2016.

\begin{table}[H]
\caption{Fractional variability amplitudes for H.E.S.S., \emph{Fermi}-LAT, Swift-XRT, Swift-UVOT and ATOM observations of PKS\,2155-304 collected in 2015 and 2016. }
\label{table:fvar2155}
\centering
\begin{tabular}{ccc}
\toprule
\textbf{Instrument} &  \textbf{ F\boldmath$_{var, 2015}$}   & \textbf{ F\boldmath$_{var, 2016}$}   \\
\midrule
  H.E.S.S.  & 0.68 $\pm$ 0.02  &  0.38 $\pm$ 0.05  \\
  LAT  & 0.53 $\pm$ 0.05    &  0.25 $\pm$ 0.05 \\
  XRT  &  1.14 $\pm$ 0.03 & 0.62 $\pm$ 0.02   \\
  UVOT (B) &  0.354 $\pm$ 0.005    & 0.234 $\pm$ 0.004   \\
  ATOM (R) & 0.322 $\pm$ 0.001    & 0.29 $\pm$ 0.01   \\
\bottomrule
\end{tabular}

 \end{table}

\subsection{Correlations}

\begin{figure}[H]
\centering
\includegraphics[width=0.3\textwidth]{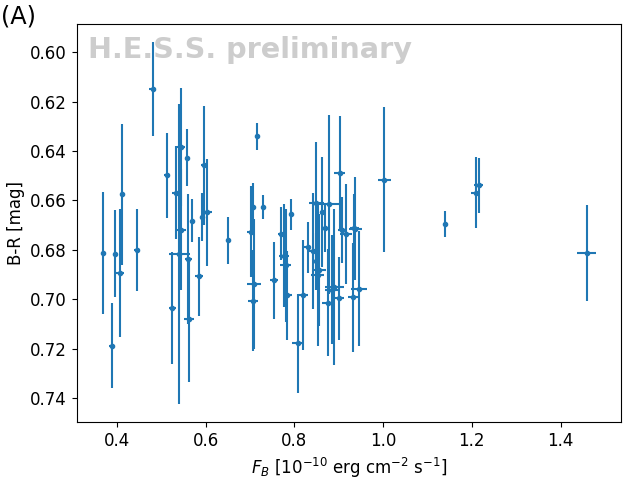}
 \includegraphics[width=0.3\textwidth]{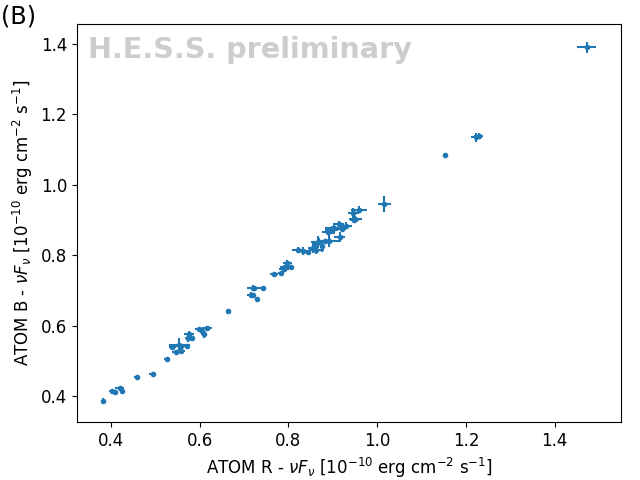}
  \includegraphics[width=0.3\textwidth]{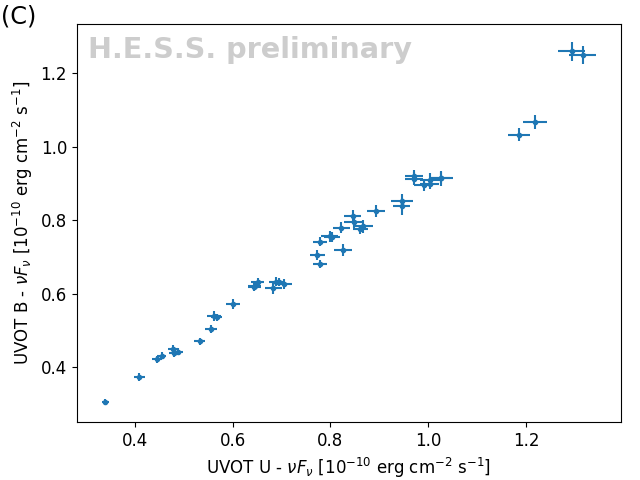}\\

  \includegraphics[width=0.3\textwidth]{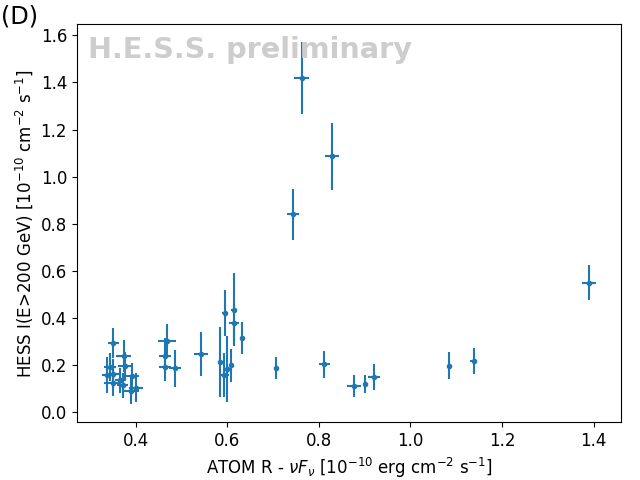}
   \includegraphics[width=0.3\textwidth]{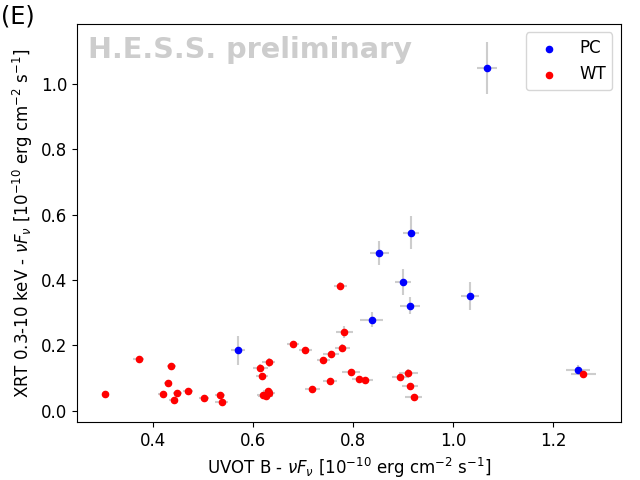}
     \includegraphics[width=0.3\textwidth]{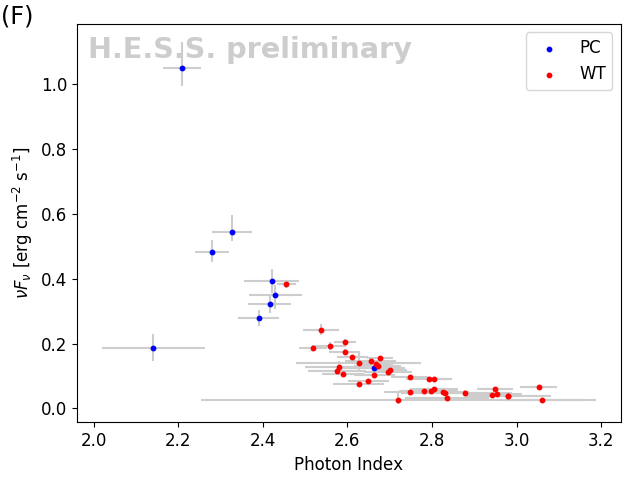}\\
\caption{Comparisons of emission observed at different wavelengths. The following panels presents: (\textbf{A}): B$-$R color vs. the B-band energy flux. (\textbf{B}): Comparison of B and R fluxes as observed with ATOM. (\textbf{C}): Comparison of optical U and B fluxes as observed with UVOT. (\textbf{D}):  VHE $\gamma$-ray photon flux of the source also as a function of the B-band energy flux. (\textbf{E}) Comparison of X-ray flux in the energy range of 0.3--10\,keV and optical flux in B band as observed with Swift-XRT and Swift-UVOT, respectively. Different colors are used to denote PC (blue) and WT (red) mode X-ray observations. (\textbf{F}) X-ray flux in the energy range of 0.3--10\,keV and corresponding photon index. Different colors are used to denote PC (blue) and WT (red) mode X-ray observations. }
\label{fig:colmag}
\end{figure}

The diagram  shown in Figure\,\ref{fig:colmag} (panel A) presents a color-magnitude diagram, i.e., B-R color vs. B band magnitude for all optical observation collected with ATOM in 2015--2016.
Since the correlation coefficient for a linear fit to the data is 0.03, no color-flux relation can be claimed in the case of 2015--2016 monitoring of PKS\,2155-304.

Panel B and C of the same plot show comparison optical (ultraviolet) fluxes of PKS\,2155-304 for the same period of observations for ATOM and UVOT observations, respectively.
For both cases, strong and significant correlation, with Pearson correlation coefficient of (0.98--0.99)~$\pm$~$<$0.01. 
Errors of the Pearson correlation coefficients were estimated as described in \cite{Wierzcholska15_48}.

The corresponding VHE $\gamma$-ray flux changes as a function of the B-band energy flux of the source are presented in the panel D of Figure\,\ref{fig:colmag}.
The correlation coefficient for the entire set of observations presented in the figure is 0.45.
In the plot, two separate trends (branches) are visible.
The lower one corresponds to the data collected during the orphan optical flare observed in 2015.
We note here also that the plot includes less data than the light curve plot since not for all night simultaneously VHE and optical observation were taken.

Almost exact simultaneous monitoring that is performed within the X-ray and optical regimes with Swift instruments, provides perfect data for X-ray-optical correction studies.
Such a comparison of the data studied in this paper is presented in Figure\,\ref{fig:colmag} (panel E).
In the case of PKS\,2155-304 both X-ray and optical emission is assigned to the low-energy bump in the broadband SED.
Thus, the correlation between both bands can be expected.
For the X-ray and optical observations of PKS\,2155-304, as shown in Figure\,\ref{fig:colmag} (panel E), the correlation coefficient is 0.45.
In the figure two branches are visible: one which follows positive correlation trend and the second, corresponding to the 2016 orphan optical flare and it is responsible for a lack of correlation.

The last panel (F) of Figure\,\ref{fig:colmag} presents a comparison of the X-ray flux and photon index.
The~trend visible in the plot,  spectra harden with the increasing flux, provides evidence for a general harder-when-brighter behavior of the blazar.
The correlation coefficient for the set of data presented is equal $-$0.75 $\pm$ 0.05.

\newpage
\paperwidth=\pdfpageheight
\paperheight=\pdfpagewidth
\pdfpageheight=\paperheight
\pdfpagewidth=\paperwidth
\newgeometry{layoutwidth=297mm,layoutheight=210 mm, left=2.7cm,right=2.7cm,top=1.8cm,bottom=1.5cm, includehead,includefoot}
\fancyheadoffset[LO,RE]{0cm}
\fancyheadoffset[RO,LE]{0cm}

\newpage
\restoregeometry
\paperwidth=\pdfpageheight
\paperheight=\pdfpagewidth
\pdfpageheight=\paperheight
\pdfpagewidth=\paperwidth
\headwidth=\textwidth

\section{Discussion and Summary}
The extragalactic sky as seen in VHE $\gamma$ rays (see e.g., \url{tevcat.uchicago.edu}) is dominated by blazars.
This population counts about 60 objects.
However, most of them have been observed very randomly and usually during the flaring state.
There are only a few extragalactic targets that can be intensively monitored in the VHE $\gamma$ rays and detected within the time of one observing run.
One of such blazars is PKS\,2155-304, which is monitored with H.E.S.S. since the beginning of its functionality.
PKS\,2155-304 is a unique blazar that can be significantly detected in the VHE regime within about 1\,h  of H.E.S.S. observations, depending on its states of activity.
In addition, this makes PKS\,2155-304 a perfect target for studies on long-term spectral and temporal variability.

During the last two decades, multiwavelength observational campaigns have become essential for studies on blazars' nature.
Despite several observations collected during this period, we have learned that observed patterns and correlations seemed to vary from object to object, and even for epoch to epoch in the case of one target (e.g., \citep[]{Abdo_2010, Rani2013, Chatterjee2013, Ahnen_501}).
This is also the case for PKS\,2155-304.
Large set of observations including VHE $\gamma$-ray data, disclosed complex behavior of the blazar (\mbox{e.g.,~\citep[]{2155_flare, 2155_chandra, 2155_corr, 2155_atom}}).

The monitoring of PKS\,2155-304 performed in 2015--2016 also presents different facets of the source.
The multiwavelength set of data revealed well correlated flaring activity observed in 2015.
The presence of correlation between optical, ultraviolet X-ray and  $\gamma$-ray regimes indicates that same physical process is responsible for the emission observed at all frequencies.
During this period of observation, the harder-when-brighter trend was also found in the X-ray data.
Such a behavior has been previously reported for several HBL type blazars including PKS\,2155-304 \citep[][]{Zhang05, Zhang06,  Kapanadze2014}.

The 2016 monitoring data exposed a completely different picture of the blazar.
This was due to the orphan flare observed only in the optical and ultraviolet range.
This optical isolated activity indicates that no obvious relation has been found between emission observed at different wavebands.
We note here that orphan optical activity has been never  reported for PKS\,2155-304 before.

Studies of the optical observations of PKS\,2155-304 collected in 2015--2016 with ATOM and Swift-UVOT reveal a strong and significant correlation between emission observed in different optical bands.
Comparison of the optical color and optical flux do not exhibit bluer-when-brighter or redder-when-brighter relation.
Previous optical observations of PKS\,2155-304 have shown different behavior of the blazar in this regime.
Lack of color-magnitude correlation according to 18 months of optical data taken with the Kanata telescope was reported \citep{Ikejiri11}.
Using the longterm ATOM monitoring performed during 792 nights in 2007–2012 the only weak bluer-when-brighter trend was noted in the entire dataset analyzed \citep{Wierzcholska_atom}.
However, the authors pointed out two separate branches in the color-magnitude diagram, with a clear correlation visible in both components.

The H.E.S.S. monitoring data together with other wavelengths observations collected in 2015--2016 again demonstrated that PKS\,2155-304  is multifaceted blazar.
It is also the argument that simultaneous, multiwavelength monitoring and deep studied on variability behavior in blazars is the only tool to search for anomalous events and disclose the nature of these sources.


\vspace{6pt}
\authorcontributions{A.W. was resposible for conceptualization and paper writing; A.W., M.Z., F.J., and S.W. provided instruments and softwares, data validation and analysis, and interpretation. }

\funding{The support of the Namibian authorities and of the University of Namibia in facilitating
the construction and operation of H.E.S.S. is gratefully acknowledged, as is the support
by the German Ministry for Education and Research (BMBF), the Max Planck Society, the
German Research Foundation (DFG), the Helmholtz Association, the Alexander von Humboldt Foundation,
the French Ministry of Higher Education, Research and Innovation, the Centre National de la
Recherche Scientifique (CNRS$/$IN2P3 and CNRS$/$INSU), the Commissariat à l’énergie atomique
et aux énergies alternatives (CEA), the U.K. Science and Technology Facilities Council (STFC),
the Knut and Alice Wallenberg Foundation, the National Science Centre, Poland grant no. 2016$/$22$/$M$/$ST9$/$00382,
the South African Department of Science and Technology and National Research Foundation, the
University of Namibia, the National Commission on Research, Science $\&$ Technology of Namibia (NCRST),
the Austrian Federal Ministry of Education, Science and Research and the Austrian Science Fund (FWF),
the Australian Research Council (ARC), the Japan Society for the Promotion of Science and by the
University of Amsterdam. We appreciate the excellent work of the technical support staff in Berlin,
Zeuthen, Heidelberg, Palaiseau, Paris, Saclay, Tübingen and in Namibia in the construction and
operation of the equipment. This work benefited from services provided by the H.E.S.S.
Virtual Organisation, supported by the national resource providers of the EGI Federation. }

\conflictsofinterest{The authors declare no conflicts of interest.}







\begin{thebibliography}{-------}
\providecommand{\natexlab}[1]{#1}

\end{thebibliography}


\begin{thebibliography}{999}
\providecommand{\natexlab}[1]{#1}

\bibitem[{Urry} and {Padovani}(1995)]{Urry95}
{Urry}, C.M.; {Padovani}, P.
\newblock {Unified Schemes for Radio-Loud Active Galactic Nuclei}.
\newblock {\em Publ.  Astron. Soc. Pac.} {\bf 1995}, {\em 107},~803,
\newblock
  doi:{\changeurlcolor{black}\href{https://doi.org/10.1086/133630}{\detokenize{10.1086/133630}}}.

\bibitem[{Begelman} \em{et~al.}(1984){Begelman}, {Blandford}, and
  {Rees}]{Begelman84}
{Begelman}, M.C.; {Blandford}, R.D.; {Rees}, M.J.
\newblock {Theory of extragalactic radio sources}.
\newblock {\em Rev. Mod.  Phys.} {\bf 1984}, {\em 56},~255--351,
\newblock
  doi:{\changeurlcolor{black}\href{https://doi.org/10.1103/RevModPhys.56.255}{\detokenize{10.1103/RevModPhys.56.255}}}.

\bibitem[{Wagner}(2009)]{Wagner09}
{Wagner}, S.
\newblock {Multiwavelength observations of Blazars}. In
\newblock  \emph{Astrophysics with All-Sky X-ray Observations}, Proceedings of the RIKEN Symposium, RIKEN, and JAXA Suzuki Umetaro Hall, RIKEN Wako, Saitama, Japan,   10--12 June 2008; p. 186.

\bibitem[{Wierzcholska} and {Siejkowski}(2015)]{Wierzcholska15L}
{Wierzcholska}, A.; {Siejkowski}, H.
\newblock {Swift/XRT view of S5 0716+714 during a flare}.
\newblock {\em Mon. Not. R. Astron. Soc. Lett.} {\bf 2015}, {\em 452},~L11--L15,
\newblock
  doi:{\changeurlcolor{black}\href{https://doi.org/10.1093/mnrasl/slv075}{\detokenize{10.1093/mnrasl/slv075}}}.

\bibitem[{Siejkowski} and {Wierzcholska}(2017)]{Siejkowski17}
{Siejkowski}, H.; {Wierzcholska}, A.
\newblock {Characterizing long-term optical, ultraviolet and X-ray variability
  in different activity states of OJ 287}.
\newblock {\em Mon. Not. R. Astron. Soc. Lett.} {\bf 2017}, {\em 468},~426--434,
\newblock
  doi:{\changeurlcolor{black}\href{https://doi.org/10.1093/mnras/stx495}{\detokenize{10.1093/mnras/stx495}}}.

\bibitem[{Wierzcholska} and {Siejkowski}(2016)]{Wierzcholska07full}
{Wierzcholska}, A.; {Siejkowski}, H.
\newblock {First hard X-ray observations of the blazar S5 0716+714 with NuSTAR
  during a multiwavelength campaign}.
\newblock {\em Mon. Not. R. Astron. Soc. Lett.} {\bf 2016}, {\em 458},~2350--2359,
\newblock
  doi:{\changeurlcolor{black}\href{https://doi.org/10.1093/mnras/stw425}{\detokenize{10.1093/mnras/stw425}}}.

\bibitem[{Maraschi} \em{et~al.}(1992){Maraschi}, {Ghisellini}, and
  {Celotti}]{Maraschi92}
{Maraschi}, L.; {Ghisellini}, G.; {Celotti}, A.
\newblock {A jet model for the gamma-ray emitting blazar 3C 279}.
\newblock {\em  Astrophys.  J.} {\bf 1992}, {\em 397},~L5--L9,
\newblock
  doi:{\changeurlcolor{black}\href{https://doi.org/10.1086/186531}{\detokenize{10.1086/186531}}}.

\bibitem[{Kirk} \em{et~al.}(1998){Kirk}, {Rieger}, and {Mastichiadis}]{Kirk98}
{Kirk}, J.G.; {Rieger}, F.M.; {Mastichiadis}, A.
\newblock {Particle acceleration and synchrotron emission in blazar jets}.
\newblock {\em Astron. Astrophys.} {\bf 1998}, {\em 333},~452--458,

\bibitem[{Dermer} \em{et~al.}(1992){Dermer}, {Schlickeiser}, and
  {Mastichiadis}]{Dermer92}
{Dermer}, C.D.; {Schlickeiser}, R.; {Mastichiadis}, A.
\newblock {High-energy gamma radiation from extragalactic radio sources}.
\newblock {\em Astron. Astrophys.} {\bf 1992}, {\em 256},~L27--L30.

\bibitem[{Sikora} \em{et~al.}(1994){Sikora}, {Begelman}, and {Rees}]{Sikora94}
{Sikora}, M.; {Begelman}, M.C.; {Rees}, M.J.
\newblock {Comptonization of diffuse ambient radiation by a relativistic jet:
  The~source of gamma rays from blazars?}
\newblock {\em  Astrophys. J.} {\bf 1994}, {\em 421},~153--162,
\newblock
  doi:{\changeurlcolor{black}\href{https://doi.org/10.1086/173633}{\detokenize{10.1086/173633}}}.

\bibitem[{Aharonian}(2000)]{Aharonian2000}
{Aharonian}, F.A.
\newblock {TeV gamma rays from BL Lac objects due to synchrotron radiation of
  extremely high energy protons}.
\newblock {\em New Astron.} {\bf 2000}, {\em 5},~377--395,
\newblock
  doi:{\changeurlcolor{black}\href{https://doi.org/10.1016/S1384-1076(00)00039-7}{\detokenize{10.1016/S1384-1076(00)00039-7}}}.

\bibitem[{Atoyan} and {Dermer}(2003)]{Atoyan2003}
{Atoyan}, A.M.; {Dermer}, C.D.
\newblock {Neutral Beams from Blazar Jets}.
\newblock {\em Astrophys. J.} {\bf 2003}, {\em 586},~79--96,
\newblock
  doi:{\changeurlcolor{black}\href{https://doi.org/10.1086/346261}{\detokenize{10.1086/346261}}}.

\bibitem[{B{\"o}ttcher} \em{et~al.}(2013){B{\"o}ttcher}, {Reimer}, {Sweeney},
  and {Prakash}]{bttcher13}
{B{\"o}ttcher}, M.; {Reimer}, A.; {Sweeney}, K.; {Prakash}, A.
\newblock {Leptonic and Hadronic Modeling of Fermi-detected Blazars}.
\newblock {\em Astrophys. J.} {\bf 2013}, {\em 768},~54,
\newblock
  doi:{\changeurlcolor{black}\href{https://doi.org/10.1088/0004-637X/768/1/54}{\detokenize{10.1088/0004-637X/768/1/54}}}.

\bibitem[{Fossati} \em{et~al.}(1998){Fossati}, {Maraschi}, {Celotti},
  {Comastri}, and {Ghisellini}]{Fossati98}
{Fossati}, G.; {Maraschi}, L.; {Celotti}, A.; {Comastri}, A.; {Ghisellini}, G.
\newblock {A unifying view of the spectral energy distributions of blazars}.
\newblock {\em Mon. Not. R. Astron. Soc. Lett.} {\bf 1998}, {\em 299},~433--448,
\newblock
  doi:{\changeurlcolor{black}\href{https://doi.org/10.1046/j.1365-8711.1998.01828.x}{\detokenize{10.1046/j.1365-8711.1998.01828.x}}}.

\bibitem[{Abdo} \em{et~al.}(2010){Abdo}, {Ackermann}, {Agudo}, {Ajello},
  {Aller}, {Aller}, {Angelakis}, {Arkharov}, {Axelsson}, {Bach}, and
  et~al.]{Abdo10}
{Abdo}, A.A.; {Ackermann}, M.; {Agudo}, I.; {Ajello}, M.; {Aller}, H.D.;
  {Aller}, M.F.; {Angelakis}, E.; {Arkharov}, A.A.; {Axelsson}, M.; {Bach}, U.;
  et~al.
\newblock {The Spectral Energy Distribution of Fermi Bright Blazars}.
\newblock {\em Astrophys. J.} {\bf 2010}, {\em 716},~30--70,
\newblock
  doi:{\changeurlcolor{black}\href{https://doi.org/10.1088/0004-637X/716/1/30}{\detokenize{10.1088/0004-637X/716/1/30}}}.

\bibitem[{Shimmins} and {Bolton}(1974)]{Shimmins74}
{Shimmins}, A.J.; {Bolton}, J.G.
\newblock {The Parkes 2700 MHz Survey (Sixth Part): Catalogue for the
  Declination zone $-$30$^\circ$ to $-$35$^\circ$}.
\newblock {\em Australian J.  Phys. Astrophys. Suppl.} {\bf
  1974}, {\em 32},~1.

\bibitem[{Hewitt} and {Burbidge}(1980)]{Hewitt80}
{Hewitt}, A.; {Burbidge}, G.
\newblock {A revised optical catalog of quasi-stellar objects}.
\newblock {\em  Astrophys. J. Suppl.  Ser.} {\bf 1980}, {\em 43},~57--158,
\newblock
  doi:{\changeurlcolor{black}\href{https://doi.org/10.1086/190667}{\detokenize{10.1086/190667}}}.

\bibitem[{Aharonian} \em{et~al.}(2007){Aharonian}, {Akhperjanian},
  {Bazer-Bachi}, {Behera}, {Beilicke}, {Benbow}, {Berge}, {Bernl{\"o}hr},
  {Boisson}, {Bolz}, {Borrel}, {Boutelier}, {Braun}, {Brion}, {Brown},
  {B{\"u}hler}, {B{\"u}sching}, {Bulik}, {Carrigan}, {Chadwick}, {Clapson},
  {Chounet}, {Coignet}, {Cornils}, {Costamante}, {Degrange}, {Dickinson},
  {Djannati-Ata{\"i}}, {Domainko}, {Drury}, {Dubus}, {Dyks}, {Egberts},
  {Emmanoulopoulos}, {Espigat}, {Farnier}, {Feinstein}, {Fiasson},
  {F{\"o}rster}, {Fontaine}, {Funk}, {Funk}, {F{\"u}{\ss}ling}, {Gallant},
  {Giebels}, {Glicenstein}, {Gl{\"u}ck}, {Goret}, {Hadjichristidis}, {Hauser},
  {Hauser}, {Heinzelmann}, {Henri}, {Hermann}, {Hinton}, {Hoffmann}, {Hofmann},
  {Holleran}, {Hoppe}, {Horns}, {Jacholkowska}, {de Jager}, {Kendziorra},
  {Kerschhaggl}, {Kh{\'e}lifi}, {Komin}, {Kosack}, {Lamanna}, {Latham}, {Le
  Gallou}, {Lemi{\`e}re}, {Lemoine-Goumard}, {Lenain}, {Lohse}, {Martin},
  {Martineau-Huynh}, {Marcowith}, {Masterson}, {Maurin}, {McComb}, {Moderski},
  {Moulin}, {de Naurois}, {Nedbal}, {Nolan}, {Olive}, {Orford}, {Osborne},
  {Ostrowski}, {Panter}, {Pedaletti}, {Pelletier}, {Petrucci}, {Pita},
  {P{\"u}hlhofer}, {Punch}, {Ranchon}, {Raubenheimer}, {Raue}, {Rayner},
  {Renaud}, {Ripken}, {Rob}, {Rolland}, {Rosier-Lees}, {Rowell}, {Rudak},
  {Ruppel}, {Sahakian}, {Santangelo}, {Saug{\'e}}, {Schlenker}, {Schlickeiser},
  {Schr{\"o}der}, {Schwanke}, {Schwarzburg}, {Schwemmer}, {Shalchi}, {Sol},
  {Spangler}, {Stawarz}, {Steenkamp}, {Stegmann}, {Superina}, {Tam},
  {Tavernet}, {Terrier}, {van Eldik}, {Vasileiadis}, {Venter}, {Vialle},
  {Vincent}, {Vivier}, {V{\"o}lk}, {Volpe}, {Wagner}, {Ward}, and
  {Zdziarski}]{2155_flare}
{Aharonian}, F.; {Akhperjanian}, A.G.; {Bazer-Bachi}, A.R.; {Behera}, B.;
  {Beilicke}, M.; {Benbow}, W.; {Berge}, D.; {Bernl{\"o}hr}, K.; {Boisson}, C.;
  {Bolz}, O.;  et al.
\newblock {An Exceptional Very High Energy Gamma-Ray Flare of PKS 2155-304}.
\newblock {\em  Astrophys.  J.} {\bf 2007}, {\em 664},~L71--L74,
\newblock
  doi:{\changeurlcolor{black}\href{https://doi.org/10.1086/520635}{\detokenize{10.1086/520635}}}.

\bibitem[{Aharonian} \em{et~al.}(2009){Aharonian}, {Akhperjanian}, {Anton},
  {Barres de Almeida}, {Bazer-Bachi}, {Becherini}, {Behera}, {Benbow},
  {Bernl{\"o}hr}, {Boisson}, {Bochow}, {Borrel}, {Brion}, {Brucker}, {Brun},
  {B{\"u}hler}, {Bulik}, {B{\"u}sching}, {Boutelier}, {Chadwick},
  {Charbonnier}, {Chaves}, {Cheesebrough}, {Chounet}, {Clapson}, {Coignet},
  {Costamante}, {Dalton}, {Daniel}, {Davids}, {Degrange}, {Deil}, {Dickinson},
  {Djannati-Ata{\"i}}, {Domainko}, {O'C.~Drury}, {Dubois}, {Dubus}, {Dyks},
  {Dyrda}, {Egberts}, {Emmanoulopoulos}, {Espigat}, {Farnier}, {Feinstein},
  {Fiasson}, {F{\"o}rster}, {Fontaine}, {F{\"u}{\ss}ling}, {Gabici}, {Gallant},
  {G{\'e}rard}, {Giebels}, {Glicenstein}, {Gl{\"u}ck}, {Goret}, {G{\"o}hring},
  {Hauser}, {Hauser}, {Heinz}, {Heinzelmann}, {Henri}, {Hermann}, {Hinton},
  {Hoffmann}, {Hofmann}, {Holleran}, {Hoppe}, {Horns}, {Jacholkowska}, {de
  Jager}, {Jahn}, {Jung}, {Katarzy{\'n}ski}, {Katz}, {Kaufmann}, {Kendziorra},
  {Kerschhaggl}, {Khangulyan}, {Kh{\'e}lifi}, {Keogh}, {Klu{\'z}niak},
  {Kneiske}, {Komin}, {Kosack}, {Lamanna}, {Lenain}, {Lohse}, {Marandon},
  {Martin}, {Martineau-Huynh}, {Marcowith}, {Maurin}, {McComb}, {Medina},
  {Moderski}, {Monard}, {Moulin}, {Naumann-Godo}, {de Naurois}, {Nedbal},
  {Nekrassov}, {Niemiec}, {Nolan}, {Ohm}, {Olive}, {de O{\~n}a Wilhelmi},
  {Orford}, {Ostrowski}, {Panter}, {Paz Arribas}, {Pedaletti}, {Pelletier},
  {Petrucci}, {Pita}, {P{\"u}hlhofer}, {Punch}, {Quirrenbach}, {Raubenheimer},
  {Raue}, {Rayner}, {Renaud}, {Rieger}, {Ripken}, {Rob}, {Rosier-Lees},
  {Rowell}, {Rudak}, {Rulten}, {Ruppel}, {Sahakian}, {Santangelo},
  {Schlickeiser}, {Sch{\"o}ck}, {Schr{\"o}der}, {Schwanke}, {Schwarzburg},
  {Schwemmer}, {Shalchi}, {Sikora}, {Skilton}, {Sol}, {Spangler}, {Stawarz},
  {Steenkamp}, {Stegmann}, {Superina}, {Szostek}, {Tam}, {Tavernet}, {Terrier},
  {Tibolla}, {Tluczykont}, {van Eldik}, {Vasileiadis}, {Venter}, {Venter},
  {Vialle}, {Vincent}, {Vivier}, {V{\"o}lk}, {Volpe}, {Wagner}, {Ward},
  {Zdziarski}, and {Zech}]{2155_chandra}
{Aharonian}, F.; {Akhperjanian}, A.G.; {Anton}, G.; {Barres de Almeida}, U.;
  {Bazer-Bachi}, A.R.; {Becherini}, Y.; {Behera}, B.; {Benbow}, W.;
  {Bernl{\"o}hr}, K.; {Boisson}, C.;  et al.
\newblock {Simultaneous multiwavelength observations of the second exceptional
  {$\gamma$}-ray flare of PKS 2155-304 in July 2006}.
\newblock {\em Astron. Astrophys.} {\bf 2009}, {\em 502},~749--770,
\newblock
  doi:{\changeurlcolor{black}\href{https://doi.org/10.1051/0004-6361/200912128}{\detokenize{10.1051/0004-6361/200912128}}}.

\bibitem[{H.E.S.S.~Collaboration} \em{et~al.}(2017){H.E.S.S.~Collaboration},
  {Abdalla}, {Abramowski}, {Aharonian}, {Ait Benkhali}, {Akhperjanian},
  {Andersson}, {Ang{\"u}ner}, {Arrieta}, {Aubert}, and et~al.]{lognorm}
{H.E.S.S.~Collaboration}; {Abdalla}, H.; {Abramowski}, A.; {Aharonian}, F.;
  {Ait Benkhali}, F.; {Akhperjanian}, A.G.; {Andersson}, T.; {Ang{\"u}ner},
  E.O.; {Arrieta}, M.; {Aubert}, P.; et~al.
\newblock {Characterizing the {$\gamma$}-ray long-term variability of PKS
  2155-304 with H.E.S.S. and Fermi-LAT}.
\newblock {\em Astron. Astrophys.} {\bf 2017}, {\em 598},~A39,
\newblock
  doi:{\changeurlcolor{black}\href{https://doi.org/10.1051/0004-6361/201629419}{\detokenize{10.1051/0004-6361/201629419}}}.

\bibitem[{Aharonian} \em{et~al.}(2006){Aharonian}, {Akhperjanian},
  {Bazer-Bachi}, {Beilicke}, {Benbow}, {Berge}, {Bernl{\"o}hr}, {Boisson},
  {Bolz}, {Borrel}, {Braun}, {Breitling}, {Brown}, {B{\"u}hler},
  {B{\"u}sching}, {Carrigan}, {Chadwick}, {Chounet}, {Cornils}, {Costamante},
  {Degrange}, {Dickinson}, {Djannati-Ata{\"i}}, {O'C.~Drury}, {Dubus},
  {Egberts}, {Emmanoulopoulos}, {Espigat}, {Feinstein}, {Ferrero}, {Fiasson},
  {Fontaine}, {Funk}, {Funk}, {Gallant}, {Giebels}, {Glicenstein}, {Goret},
  {Hadjichristidis}, {Hauser}, {Hauser}, {Heinzelmann}, {Henri}, {Hermann},
  {Hinton}, {Hofmann}, {Holleran}, {Horns}, {Jacholkowska}, {de Jager},
  {Kh{\'e}lifi}, {Komin}, {Konopelko}, {Kosack}, {Latham}, {Le Gallou},
  {Lemi{\`e}re}, {Lemoine-Goumard}, {Lohse}, {Martin}, {Martineau-Huynh},
  {Marcowith}, {Masterson}, {McComb}, {de Naurois}, {Nedbal}, {Nolan},
  {Noutsos}, {Orford}, {Osborne}, {Ouchrif}, {Panter}, {Pelletier}, {Pita},
  {P{\"u}hlhofer}, {Punch}, {Raubenheimer}, {Raue}, {Rayner}, {Reimer},
  {Reimer}, {Ripken}, {Rob}, {Rolland}, {Rowell}, {Sahakian}, {Saug{\'e}},
  {Schlenker}, {Schlickeiser}, {Schwanke}, {Sol}, {Spangler}, {Spanier},
  {Steenkamp}, {Stegmann}, {Superina}, {Tavernet}, {Terrier}, {Th{\'e}oret},
  {Tluczykont}, {van Eldik}, {Vasileiadis}, {Venter}, {Vincent}, {V{\"o}lk},
  {Wagner}, and {Ward}]{Aharonian2006_crab}
{Aharonian}, F.; {Akhperjanian}, A.G.; {Bazer-Bachi}, A.R.; {Beilicke}, M.;
  {Benbow}, W.; {Berge}, D.; {Bernl{\"o}hr}, K.; {Boisson}, C.; {Bolz}, O.;
  {Borrel}, V.;  et al.
\newblock {Observations of the Crab nebula with HESS}.
\newblock {\em Astron. Astrophys.} {\bf 2006}, {\em 457},~899--915,
\newblock
  doi:{\changeurlcolor{black}\href{https://doi.org/10.1051/0004-6361:20065351}{\detokenize{10.1051/0004-6361:20065351}}}.

\bibitem[{Parsons} and {Hinton}(2014)]{Parsons14}
{Parsons}, R.D.; {Hinton}, J.A.
\newblock {A Monte Carlo template based analysis for air-Cherenkov arrays}.
\newblock {\em Astropart. Phys.} {\bf 2014}, {\em 56},~26--34,
\newblock
  doi:{\changeurlcolor{black}\href{https://doi.org/10.1016/j.astropartphys.2014.03.002}{\detokenize{10.1016/j.astropartphys.2014.03.002}}}.

\bibitem[{de Naurois} and {Rolland}(2009)]{Naurois09}
{de Naurois}, M.; {Rolland}, L.
\newblock {A high performance likelihood reconstruction of {$\gamma$}-rays for
  imaging atmospheric Cherenkov telescopes}.
\newblock {\em Astropart. Phys.} {\bf 2009}, {\em 32},~231--252,
\newblock
  doi:{\changeurlcolor{black}\href{https://doi.org/10.1016/j.astropartphys.2009.09.001}{\detokenize{10.1016/j.astropartphys.2009.09.001}}}.

\bibitem[{Atwood} \em{et~al.}(2013){Atwood}, {Albert}, {Baldini}, {Tinivella},
  {Bregeon}, {Pesce-Rollins}, {Sgr{\`o}}, {Bruel}, {Charles}, {Drlica-Wagner},
  {Franckowiak}, {Jogler}, {Rochester}, {Usher}, {Wood}, {Cohen-Tanugi}, and
  {S.~Zimmer for the Fermi-LAT Collaboration}]{Atwood13}
{Atwood}, W.; {Albert}, A.; {Baldini}, L.; {Tinivella}, M.; {Bregeon}, J.;
  {Pesce-Rollins}, M.; {Sgr{\`o}}, C.; {Bruel}, P.; {Charles},~E.;
  {Drlica-Wagner}, A.;  et al.
\newblock {Pass 8: Toward the Full Realization of the Fermi-LAT Scientific
  Potential}.
\newblock {\em arXiv} {\bf 2013},
  arXiv:1303.3514.

\bibitem[{Mattox} \em{et~al.}(1996){Mattox}, {Bertsch}, {Chiang}, {Dingus},
  {Digel}, {Esposito}, {Fierro}, {Hartman}, {Hunter}, {Kanbach}, {Kniffen},
  {Lin}, {Macomb}, {Mayer-Hasselwander}, {Michelson}, {von Montigny},
  {Mukherjee}, {Nolan}, {Ramanamurthy}, {Schneid}, {Sreekumar}, {Thompson}, and
  {Willis}]{Mattox96}
{Mattox}, J.R.; {Bertsch}, D.L.; {Chiang}, J.; {Dingus}, B.L.; {Digel}, S.W.;
  {Esposito}, J.A.; {Fierro}, J.M.; {Hartman}, R.C.; {Hunter}, S.D.; {Kanbach},
  G.;  et al.
\newblock {Likelihood Analysis of EGRET Data}.
\newblock {\em Astrophys. J.} {\bf 1996}, {\em 461},~396,
\newblock
  doi:{\changeurlcolor{black}\href{https://doi.org/10.1086/177068}{\detokenize{10.1086/177068}}}.

\bibitem[{Kalberla} \em{et~al.}(2005){Kalberla}, {Burton}, {Hartmann}, {Arnal},
  {Bajaja}, {Morras}, and {P{\"o}ppel}]{Kalberla2005}
{Kalberla}, P.M.W.; {Burton}, W.B.; {Hartmann}, D.; {Arnal}, E.M.; {Bajaja},
  E.; {Morras}, R.; {P{\"o}ppel}, W.G.L.
\newblock {The Leiden/Argentine/Bonn (LAB) Survey of Galactic HI. Final data
  release of the combined LDS and IAR surveys with improved stray-radiation
  corrections}.
\newblock {\em Astron. Astrophys.} {\bf 2005}, {\em 440},~775--782,
\newblock
  doi:{\changeurlcolor{black}\href{https://doi.org/10.1051/0004-6361:20041864}{\detokenize{10.1051/0004-6361:20041864}}}.

\bibitem[{Poole} \em{et~al.}(2008){Poole}, {Breeveld}, {Page}, {Landsman},
  {Holland}, {Roming}, {Kuin}, {Brown}, {Gronwall}, {Hunsberger}, {Koch},
  {Mason}, {Schady}, {vanden Berk}, {Blustin}, {Boyd}, {Broos}, {Carter},
  {Chester}, {Cucchiara}, {Hancock}, {Huckle}, {Immler}, {Ivanushkina},
  {Kennedy}, {Marshall}, {Morgan}, {Pandey}, {de Pasquale}, {Smith}, and
  {Still}]{Poole08}
{Poole}, T.S.; {Breeveld}, A.A.; {Page}, M.J.; {Landsman}, W.; {Holland}, S.T.;
  {Roming}, P.; {Kuin}, N.P.M.; {Brown}, P.J.; {Gronwall}, C.; {Hunsberger},
  S.;  et al.
\newblock {Photometric calibration of the Swift ultraviolet/optical telescope}.
\newblock {\em Mon. Not. R. Astron. Soc. Lett.} {\bf 2008}, {\em 383},~627--645,
\newblock
  doi:{\changeurlcolor{black}\href{https://doi.org/10.1111/j.1365-2966.2007.12563.x}{\detokenize{10.1111/j.1365-2966.2007.12563.x}}}.

\bibitem[{Schlegel} \em{et~al.}(1998){Schlegel}, {Finkbeiner}, and
  {Davis}]{Schlegel98}
{Schlegel}, D.J.; {Finkbeiner}, D.P.; {Davis}, M.
\newblock {Maps of Dust Infrared Emission for Use in Estimation of Reddening
  and Cosmic Microwave Background Radiation Foregrounds}.
\newblock {\em Astrophys. J.} {\bf 1998}, {\em 500},~525--553,
\newblock
  doi:{\changeurlcolor{black}\href{https://doi.org/10.1086/305772}{\detokenize{10.1086/305772}}}.

\bibitem[{Schlafly} and {Finkbeiner}(2011)]{Schlafly}
{Schlafly}, E.F.; {Finkbeiner}, D.P.
\newblock {Measuring Reddening with Sloan Digital Sky Survey Stellar Spectra
  and Recalibrating SFD}.
\newblock {\em Astrophys. J.} {\bf 2011}, {\em 737},~103,
\newblock
  doi:{\changeurlcolor{black}\href{https://doi.org/10.1088/0004-637X/737/2/103}{\detokenize{10.1088/0004-637X/737/2/103}}}.

\bibitem[{Hauser} \em{et~al.}(2004){Hauser}, {M{\"o}llenhoff}, {P{\"u}hlhofer},
  {Wagner}, {Hagen}, and {Knoll}]{Hauser}
{Hauser}, M.; {M{\"o}llenhoff}, C.; {P{\"u}hlhofer}, G.; {Wagner}, S.J.;
  {Hagen}, H.J.; {Knoll}, M.
\newblock {ATOM---An Automatic Telescope for Optical Monitoring}.
\newblock {\em Astron. Nachr.} {\bf 2004}, {\em 325},~659--659,
\newblock
  doi:{\changeurlcolor{black}\href{https://doi.org/10.1002/asna.200410305}{\detokenize{10.1002/asna.200410305}}}.

\bibitem[{Fukugita} \em{et~al.}(1995){Fukugita}, {Shimasaku}, and
  {Ichikawa}]{Fukugita95}
{Fukugita}, M.; {Shimasaku}, K.; {Ichikawa}, T.
\newblock {Galaxy Colors in Various Photometric Band Systems}.
\newblock {\em Publ.  Astron. Soc. Pac.} {\bf 1995}, {\em 107},~945,
\newblock
  doi:{\changeurlcolor{black}\href{https://doi.org/10.1086/133643}{\detokenize{10.1086/133643}}}.

\bibitem[{Falomo} \em{et~al.}(1991){Falomo}, {Giraud}, {Maraschi}, {Melnick},
  {Tanzi}, and {Treves}]{Falomo91}
{Falomo}, R.; {Giraud}, E.; {Maraschi}, L.; {Melnick}, J.; {Tanzi}, E.G.;
  {Treves}, A.
\newblock {On the nebulosity surrounding the BL Lacertae object PKS 2155-304}.
\newblock {\em  Astrophys.  J.} {\bf 1991}, {\em 380},~L67--L69,
\newblock
  doi:{\changeurlcolor{black}\href{https://doi.org/10.1086/186175}{\detokenize{10.1086/186175}}}.

\bibitem[{Vaughan} \em{et~al.}(2003){Vaughan}, {Edelson}, {Warwick}, and
  {Uttley}]{Vaughan2003}
{Vaughan}, S.; {Edelson}, R.; {Warwick}, R.S.; {Uttley}, P.
\newblock {On characterizing the variability properties of X-ray light curves
  from active galaxies}.
\newblock {\em Mon. Not. R. Astron. Soc. Lett.} {\bf 2003}, {\em 345},~1271--1284,
\newblock
  doi:{\changeurlcolor{black}\href{https://doi.org/10.1046/j.1365-2966.2003.07042.x}{\detokenize{10.1046/j.1365-2966.2003.07042.x}}}.

\bibitem[{Wierzcholska}(2015)]{Wierzcholska15_48}
{Wierzcholska}, A.
\newblock {Nine years of multi-frequency monitoring of the blazar PKS 0048-097:
  spectral and temporal variability}.
\newblock {\em Astron. Astrophys.} {\bf 2015}, {\em 580},~A104,
\newblock
  doi:{\changeurlcolor{black}\href{https://doi.org/10.1051/0004-6361/201525649}{\detokenize{10.1051/0004-6361/201525649}}}.

\bibitem[{Abdo} \em{et~al.}(2010){Abdo}, {Ackermann}, {Ajello}, {Axelsson},
  {Baldini}, {Ballet}, {Barbiellini}, {Bastieri}, {Baughman}, {Bechtol}, and
  et~al.]{Abdo_2010}
{Abdo}, A.A.; {Ackermann}, M.; {Ajello}, M.; {Axelsson}, M.; {Baldini}, L.;
  {Ballet}, J.; {Barbiellini}, G.; {Bastieri}, D.; {Baughman}, B.M.; {Bechtol},
  K.; et~al.
\newblock {A change in the optical polarization associated with a
  {$\gamma$}-ray flare in the blazar 3C279}.
\newblock {\em Nature} {\bf 2010}, {\em 463},~919--923,
\newblock
  doi:{\changeurlcolor{black}\href{https://doi.org/10.1038/nature08841}{\detokenize{10.1038/nature08841}}}.

\bibitem[{Rani} \em{et~al.}(2013){Rani}, {Krichbaum}, {Fuhrmann},
  {B{\"o}ttcher}, {Lott}, {Aller}, {Aller}, {Angelakis}, {Bach}, {Bastieri},
  {Falcone}, {Fukazawa}, {Gabanyi}, {Gupta}, {Gurwell}, {Itoh}, {Kawabata},
  {Krips}, {L{\"a}hteenm{\"a}ki}, {Liu}, {Marchili}, {Max-Moerbeck},
  {Nestoras}, {Nieppola}, {Quintana-Lacaci}, {Readhead}, {Richards}, {Sasada},
  {Sievers}, {Sokolovsky}, {Stroh}, {Tammi}, {Tornikoski}, {Uemura},
  {Ungerechts}, {Urano}, and {Zensus}]{Rani2013}
{Rani}, B.; {Krichbaum}, T.P.; {Fuhrmann}, L.; {B{\"o}ttcher}, M.; {Lott}, B.;
  {Aller}, H.D.; {Aller}, M.F.; {Angelakis}, E.; {Bach},~U.; {Bastieri}, D.; et al.
\newblock {Radio to gamma-ray variability study of blazar S5 0716+714}.
\newblock {\em Astron. Astrophys.} {\bf 2013}, {\em 552},~A11,
\newblock
  doi:{\changeurlcolor{black}\href{https://doi.org/10.1051/0004-6361/201321058}{\detokenize{10.1051/0004-6361/201321058}}}.

\bibitem[{Chatterjee} \em{et~al.}(2013){Chatterjee}, {Nalewajko}, and
  {Myers}]{Chatterjee2013}
{Chatterjee}, R.; {Nalewajko}, K.; {Myers}, A.D.
\newblock {Implications of the Anomalous Outburst in the Blazar PKS 0208-512}.
\newblock {\em  Astrophys.  J.} {\bf 2013}, {\em 771},~L25,
\newblock
  doi:{\changeurlcolor{black}\href{https://doi.org/10.1088/2041-8205/771/2/L25}{\detokenize{10.1088/2041-8205/771/2/L25}}}.

\bibitem[{Ahnen} \em{et~al.}(2018){Ahnen}, {Ansoldi}, {Antonelli}, {Arcaro},
  {Babi{\'c}}, {Banerjee}, {Bangale}, {Barres de Almeida}, {Barrio}, {Becerra
  Gonz{\'a}lez}, and et~al.]{Ahnen_501}
{Ahnen}, M.L.; {Ansoldi}, S.; {Antonelli}, L.A.; {Arcaro}, C.; {Babi{\'c}}, A.;
  {Banerjee}, B.; {Bangale}, P.; {Barres de Almeida},~U.; {Barrio}, J.A.;
  {Becerra Gonz{\'a}lez}, J.; et~al.
\newblock {The extreme HBL behaviour of Markarian 501 during 2012}.
\newblock {\em arXiv} {\bf 2018},
  arXiv:1808.04300.

\bibitem[{Aharonian} \em{et~al.}(2009){Aharonian}, {Akhperjanian}, {Anton},
  {Barres de Almeida}, {Bazer-Bachi}, {Becherini}, {Behera}, {Bernl{\"o}hr},
  {Boisson}, {Bochow}, and et~al.]{2155_corr}
{Aharonian}, F.; {Akhperjanian}, A.G.; {Anton}, G.; {Barres de Almeida}, U.;
  {Bazer-Bachi}, A.R.; {Becherini}, Y.; {Behera}, B.; {Bernl{\"o}hr}, K.;
  {Boisson}, C.; {Bochow}, A.; et~al.
\newblock {Simultaneous Observations of PKS 2155-304 with HESS, Fermi, RXTE,
  and Atom: Spectral Energy Distributions and Variability in a Low State}.
\newblock {\em  Astrophys.  J.} {\bf 2009}, {\em 696},~L150--L155,
\newblock
  doi:{\changeurlcolor{black}\href{https://doi.org/10.1088/0004-637X/696/2/L150}{\detokenize{10.1088/0004-637X/696/2/L150}}}.

\bibitem[{H.E.S.S.~Collaboration} \em{et~al.}(2014){H.E.S.S.~Collaboration},
  {Abramowski}, {Aharonian}, {Ait Benkhali}, {Akhperjanian}, {Ang{\"u}ner},
  {Backes}, {Balenderan}, {Balzer}, {Barnacka}, and et~al.]{2155_atom}
{H.E.S.S.~Collaboration}; {Abramowski}, A.; {Aharonian}, F.; {Ait Benkhali},
  F.; {Akhperjanian}, A.G.; {Ang{\"u}ner},~E.O.; {Backes}, M.; {Balenderan},
  S.; {Balzer}, A.; {Barnacka}, A.; et~al.
\newblock {Long-term monitoring of PKS 2155-304 with ATOM and H.E.S.S.:
  investigation of optical/{$\gamma$}-ray correlations in different spectral
  states}.
\newblock {\em Astron. Astrophys.} {\bf 2014}, {\em 571},~A39,
\newblock
  doi:{\changeurlcolor{black}\href{https://doi.org/10.1051/0004-6361/201424142}{\detokenize{10.1051/0004-6361/201424142}}}.

\bibitem[{Zhang} \em{et~al.}(2005){Zhang}, {Treves}, {Celotti}, {Qin}, and
  {Bai}]{Zhang05}
{Zhang}, Y.H.; {Treves}, A.; {Celotti}, A.; {Qin}, Y.P.; {Bai}, J.M.
\newblock {XMM-Newton View of PKS 2155-304: Characterizing the X-Ray
  Variability Properties with EPIC pn}.
\newblock {\em Astrophys. J.} {\bf 2005}, {\em 629},~686--699,
\newblock
  doi:{\changeurlcolor{black}\href{https://doi.org/10.1086/431719}{\detokenize{10.1086/431719}}}.

\bibitem[{Zhang} \em{et~al.}(2006){Zhang}, {Bai}, {Zhang}, {Treves},
  {Maraschi}, and {Celotti}]{Zhang06}
{Zhang}, Y.H.; {Bai}, J.M.; {Zhang}, S.N.; {Treves}, A.; {Maraschi}, L.;
  {Celotti}, A.
\newblock {Multiwavelength Observations of the BL Lacertae Object PKS 2155-304
  with XMM-Newton}.
\newblock {\em Astrophys. J.} {\bf 2006}, {\em 651},~782--790,
\newblock
  doi:{\changeurlcolor{black}\href{https://doi.org/10.1086/507601}{\detokenize{10.1086/507601}}}.

\bibitem[{Kapanadze} \em{et~al.}(2014){Kapanadze}, {Romano}, {Vercellone}, and
  {Kapanadze}]{Kapanadze2014}
{Kapanadze}, B.; {Romano}, P.; {Vercellone}, S.; {Kapanadze}, S.
\newblock {The X-ray behaviour of the high-energy peaked BL Lacertae source PKS
  2155-304 in the 0.3--10 keV band}.
\newblock {\em Mon. Not. R. Astron. Soc. Lett.} {\bf 2014}, {\em 444},~1077--1094,
\newblock
  doi:{\changeurlcolor{black}\href{https://doi.org/10.1093/mnras/stu1504}{\detokenize{10.1093/mnras/stu1504}}}.

\bibitem[{Ikejiri} \em{et~al.}(2011){Ikejiri}, {Uemura}, {Sasada}, {Ito},
  {Yamanaka}, {Sakimoto}, {Arai}, {Fukazawa}, {Ohsugi}, {Kawabata}, {Yoshida},
  {Sato}, and {Kino}]{Ikejiri11}
{Ikejiri}, Y.; {Uemura}, M.; {Sasada}, M.; {Ito}, R.; {Yamanaka}, M.;
  {Sakimoto}, K.; {Arai}, A.; {Fukazawa}, Y.; {Ohsugi},~T.; {Kawabata}, K.S.; et al.
\newblock {Photopolarimetric Monitoring of Blazars in the Optical and
  Near-Infrared Bands with the Kanata Telescope. I. Correlations between Flux,
  Color, and Polarization}.
\newblock {\em Publ. Astron. Soc. Jpn.} {\bf 2011}, {\em 63},~639--675,
\newblock
  doi:{\changeurlcolor{black}\href{https://doi.org/10.1093/pasj/63.3.327}{\detokenize{10.1093/pasj/63.3.327}}}.

\bibitem[{Wierzcholska} \em{et~al.}(2015){Wierzcholska}, {Ostrowski},
  {Stawarz}, {Wagner}, and {Hauser}]{Wierzcholska_atom}
{Wierzcholska}, A.; {Ostrowski}, M.; {Stawarz}, {\L}.; {Wagner}, S.; {Hauser},
  M.
\newblock {Longterm optical monitoring of bright BL Lacertae objects with ATOM:
  Spectral variability and multiwavelength correlations}.
\newblock {\em Astron. Astrophys.} {\bf 2015}, {\em 573},~A69,
\newblock
  doi:{\changeurlcolor{black}\href{https://doi.org/10.1051/0004-6361/201423967}{\detokenize{10.1051/0004-6361/201423967}}}.

\end{thebibliography}
\reftitle{References}


\end{document}